\documentclass[twocolumn,floats,pra]{revtex4}

\newcommand {\be}{\begin{equation}}
\newcommand {\ee} {\end{equation}}
\newcommand {\bea}{\begin{eqnarray}}
\newcommand {\eea} {\end{eqnarray}}
\newcommand{\non}{\nonumber}

\begin{document}


\title{Quasiparticle spin from adiabatic transport in quantum Hall trial wavefunctions}
\author{N. Read}
\affiliation{Department of Physics, Yale
University, P.O. Box 208120, New Haven, CT 06520-8120, USA}
\date{July 19, 2008}

\begin{abstract}
Quasiparticle spin (in the spacetime sense) couples to curvature of space. Here this fact is used to calculate the spin of quasiholes in trial quantum Hall states by adiabatically dragging them around on a sphere, for trial states given by conformal blocks in some conformal field theory. The spin is found to agree with the conformal weight of the corresponding field. The result completes a recent argument that constructions using blocks from non-unitary theories that contain negative quantum dimensions produce contradictions that prevent them from describing topological (gapped) phases of matter.
\end{abstract}

\pacs{PACS numbers: } 

\maketitle


Recently, increased attention has been paid to non-Abelian quantum Hall (QH) states. Much of the theory underlying these is based on the use of conformal blocks taken from conformal field theories as trial wavefunctions for a many-particle state. In particular, in a recent paper \cite{read08b}, it was shown that if the perturbed CFT is in a massive phase, then the adiabatic statistics of the quasiholes can be calculated and agrees with the monodromy of the conformal blocks. We also used similar arguments to obtain the Hall viscosity from adiabatic variation of the system geometry. Finally, arguments were given that when the underlying CFT is non-unitary, or more particularly when it contains negative quantum dimensions, then the resulting topological phase does not obey the positivity of quantum dimensions that follows from standard quantum mechanics, and so the trial wavefunctions cannot represent a topological phase. If a ``special'' Hamiltonian for the trial states are exact zero-energy eigenstates exists for such functions, then it apparently must be gapless. The argument went through for some examples, but was not complete for a general such construction because the so-called twist for the quasiparticles, which must be calculated adiabatically, was not available. The twist $\theta_\alpha$ associated with each quasiparticle type $\alpha$ is related to the spin $s_\alpha$ by $\theta_\alpha=e^{2\pi is_\alpha}$, because the twist essentially corresponds to rotating the quasiparticle by $2\pi$ about its position.

In this note, we return to the issue of quasiparticle twist by performing an adiabatic calculation of the spin of a quasihole. Instead of the adiabatic transport in the space of metrics, as proposed for this purpose in Ref.\ \cite{read08b}, we simply transport the position of the quasihole, as in \cite{asw}, but in the presence of curvature of the space on which the QH state
lives. As is well-known, in $2+1$ dimensions, a particle with spin (in the spacetime sense) experiences curvature in a similar way as a charged particle experiences magnetic field: adiabatic transport around a closed loop picks up a Berry phase related to the flux of curvature though the loop. By using arguments similar to those of the previous paper, the quasihole spin can be extracted on the basis of an assumption that corresponds to being in a topological phase of matter. Once again, the result is that the spin, and hence the twist, are the same as those in the underlying CFT. This allows the completion of the argument that non-unitary theories in which some quantum dimensions are negative (which includes all those in which some scaling dimensions are negative) do not give rise to sensible topological phases of matter.

Let us will begin to address the problem with a simple-minded approach. Suppose that we have some topological phase of matter, on a sphere. Suppose further that there are just two quasiparticles over the ground state, one of type $\alpha$, one of type $\alpha^*$, the ``anti-particle'' of $\alpha$. That is, these two can fuse to the identity quasiparticle type, which means that states with only these two quasiparticles do exist. The two quasiparticles are assumed to be well separated compared with the microscopic correlation length \cite{read08b}, which is independent of system size, and the latter is assumed to be large. If the separation of the two is much less than $R$, then in their neighborhood the space can be viewed as flat. Adiabatically transporting type $\alpha$ around type $\alpha^*$, with say $\alpha^*$ at the north pole, counterclockwise in the local coordinates, produces a well-defined exchange phase factor $c_{\alpha,\alpha^*}^0c_{\alpha^*,\alpha}^0$ (the notation is not important, but anyway was defined in Ref.\ \cite{read08b}); here we neglect charge and background magnetic field for the moment. If the phase of matter corresponds to a chiral conformal field theory, this phase factor is $e^{-4\pi i h_\alpha}$, where $h_\alpha=h_{\alpha^*}$ are the conformal weights associated with the corresponding fields, and can be read off from the two-point function $1/z^{2h_\alpha}$.

But now suppose that the separation of the quasiparticles is increased, and the exchange is considered always with the separation fixed in magnitude. For very large separation, type $\alpha$ can be viewed as encircling the south pole. As the radius of the circle about the south pole goes to zero, the phase factor should approach $1$. There is no conflict with the previous paragraph, provided that the quasiparticles carry non-zero spin. The spin couples to curvature, and gives a phase $s_\alpha$ times the integral of the scalar (or Ricci) curvature $\cal R$ enclosed by the path. The curvature integrated over the sphere is $4\pi$ (by the Gauss-Bonnet theorem), so the phase factor for a path enclosing the full sphere would be $e^{4\pi i s_\alpha}$. Thus, for our problem, adding the result for  small separation to that due to curvature times spin, we require that $e^{4\pi i s_\alpha}c_{\alpha,\alpha^*}^0c_{\alpha^*,\alpha}^0=1$. That is, using $\theta_\alpha=e^{2\pi i s_\alpha}$, we have shown that %
\be%
\theta_\alpha^{-2}=c_{\alpha,\alpha^*}^0c_{\alpha^*,\alpha}^0,\ee%
which determines the twist up to a sign. (This relation is known from topological considerations; see e.g.\ Ref.\ \cite{read08b} and references therein.) Notice that we did not assume the statistics is Abelian, and in general this result is not a spin-statistics theorem (though it is close to it in the Abelian case). Non-Abelian statistics can show up when a sufficient number of quasiparticles are present. But if many quasiparticles other than the one of type $\alpha$ that we transport are present, the others must fuse to type $\alpha^*$, and the same argument goes through, even if the other quasiparticles are not all close to the north pole, and even if some of the quasiparticles are non-Abelian. We note that in the influential paper Ref.\ \cite{tw}, the coupling of the spins of the Abelian anyons to curvature is omitted, resulting in a rather different relation for anyons on a sphere.

To obtain the spin uniquely, we will calculate directly the Berry connection for transporting the quasiparticle around a small loop. This will be considered here for trial wavefunctions given by conformal blocks, which will be formulated on the sphere with uniform curvature and magnetic field. We begin with the formalism for this.

We take the sphere to have the standard rotationally-invariant metric, and rotationally-invariant field strength. The formulation of QH states on the sphere is known \cite{hald83}. The particle wavefunctions can be written in terms of homogeneous or spinor coordinates $(u_i,v_i)$ (with $|u_i|^2+|v_i|^2=1$) for the $i$th particle. Alternatively, by stereographic projection of the sphere to the plane, we can introduce the coordinates $z_i=2R v_i/u_i$ \cite{rr1}. The wavefunctions for basis states in the lowest Landau level, in the symmetric gauge, are \cite{rr1}%
\be%
z^m/\left(1+|z|^2/(4R^2)\right)^{N_\phi/2+1},\ee%
where $m$ and $N_\phi$ are non-negative integers.
Here we assume that the norm-square of these functions is to be calculated by taking the mod-square of the function, and integrating with over $z$ with measure $dzd\bar{z}$. Then the basis states are normalizable only for $m=0$, $1$, \ldots, $N_\phi$. $N_\phi$ is the number of flux piericing the sphere, and if $z=0$ is regarded as the north pole of the sphere, then $N_\phi/2-m$ is the $3$-component of angular momentum.

A little more depth will be useful later in this paper. First, the $+1$ in the exponent is due to the definition of the integration measure as $dzd\bar{z}$. It came from the rotationally invariant measure on the sphere, which is $dzd\bar{z}/\left(1+|z|^2/(4R^2)\right)^2$. Recall that if a general coordinate system with coordinates $x^a$ is used for a manifold, and a metric $g_{ab}$, defined by the squared line element $ds^2=\sum_{a,b} g_{ab}dx^a dx^b$ is given, then the covariant integration measure, which weights a box of sides $dx^a$ by its volume computed using the metric is $\sqrt{|g|}\prod_a dx_a$, where $|g|$ stands for $|\det g_{ab}|$. For the present coordinates for the sphere, the line element is%
\be%
ds^2=\frac{|dz|^2}{\left(1+|z|^2/(4R^2)\right)^2} .\ee%
Hence, this factor should be removed from the wavefunctions if we wish to take the usual view of them as scalar functions under (time-independent) coordinate transformations. This view, which will be taken from here on, is used in writing down the Landau Hamiltonian, which is a gauge-covariant version of the Laplacian on the sphere, acting on scalar functions.

Second, the number of remaining factors $1/\left(1+|z|^2/(4R^2)\right)^{1/2}$ is equal to the number of flux $N_\phi$. For the highest allowed $m$ value, the basis function (viewed as we said as a scalar function) approaches a constant in magnitude as $|z|\to\infty$, but winds in phase by $2\pi N_\phi$ as $\arg z$ varies. On the sphere, this basis function should go to a constant as $z\to\infty$, which corresponds to the south pole. The apparent conflict (and similar ones for the other basis states) in the phase is resolved because a distinct coordinate patch is needed to cover the south pole. The patch for the south pole is represented by coordinate $z'$, and clearly from the spinor coordinates $z'=4R^2/z$ for $z'$, $z\neq 0$, $\infty$. Thus the two patches overlap in an annulus that can be contracted to a circle represented by, for example, any circle of $z$ constant. A U$(1)$ gauge transformation is performed in going from one patch to the other, and the essential information in the transformation is how many times it winds on making a circuit of the circle. This integer defines the first Chern class of the {\em bundle} whose sections are defined by the ``functions'' on the two coordinate patches that agree under this gauge transformation. Thus the basis states, and the space of functions they span, are not actually functions, but sections of a bundle. The Chern class is simply the number of magnetic flux involved (compare with the notion of ``Dirac string''). The particular space of sections we call the LLL can be called holomorphic sections, as they are holomorphic in the above patches, up to the ubiquitous factor.

We already see how to read off the number of flux $N_\phi$ in the one-particle problem from a given set of basis functions. In the many-particle situation, we should use adiabatic transport of particles or quasiparticles, and calculate the curl of the Berry connection to obtain field strength. We can do this in the single-particle problem, so that the later calculation is reduced to counting. A coherent-state wavepacket for the charged particle in the LLL can be written as $(u\bar{\alpha}+v\bar{\beta})^{N_\phi}$ \cite{hald83}, which in terms of $z$ and $w=2R\beta/\alpha$ becomes %
\be%
\frac{(1+z\bar{w}/(4R^2))^{N_\phi}}{\left(1+|z|^2/(4R^2)\right)^{N_\phi/2}
\left(1+
|w|^2/(4R^2)\right)^{N_\phi/2}}.\ee%
This is peaked near $z=w$, and reduces to a Gaussian packet as $N_\phi\to\infty$ with $N_\phi/R^2$ fixed. In addition, it is normalized, up to a factor independent of $w$. It is anti-holomorphic in $w$, except for the normalizing factor in the denominator. This makes it possible to apply the approach to the Berry phase (treating $w$ as parameter) of Ref.\ \cite{read08b}. The Berry connection is $A_{\bar{w}}=iN_\phi w/\left[(8R^2)\left(1+
|w|^2/(4R^2)\right)\right]$, $A_w=\overline{A_{\bar{w}}}$. Then the curl is $\partial_w A_{\bar{w}}-\partial_{\bar{w}}A_w=i N_\phi/\left[4R^2\left(1+
|w|^2/(4R^2)\right)\right]$. Using the metric, the number of flux per unit area is then $N_\phi/2R^2$, as would be expected.

Now we turn to many-particle wavefunctions, beginning with the Laughlin states, and progressing to more general states afterwards. As in some other adiabatic calculations \cite{read08b}, the latter step is in some ways straightforward, and most of the complications are due to the background magnetic field/charge density arising from the charge part of the states, which is already present in the Laughlin case. (The result for the Laughlin states may be known, though we are not aware of a reference.)

The holomorphic part of the Laughlin state in the disk geometry consists solely of factors $z_i-z_j$. On the sphere, each of these becomes $u_iv_j-v_iu_j$ which are manifestly rotationally invariant \cite{hald83}, and in terms of the coordinates $z_i$ become (again in the symmetric gauge)%
\be%
\frac{z_i-z_j}{\left(1+\frac{|z_i|^2}{4R^2}\right)^{1/2}\left(1+\frac{|z_j|^2}
{4R^2}\right)^{1/2}},\ee%
up to overall factors of $R$, which are unimportant. Then the Laughlin state of exponent $Q$ with $N$ particles at positions $z_i$, $i=1$, \ldots, $N$, and $n$ quasiholes at arbitrary positions $w_k$, $k=1$, \ldots, $n$, becomes%
\bea%
&&
\prod_{i,k}(z_i-w_k)\cdot\prod_{i<j}(z_i-z_j)^Q\\
&&{}\cdot
\prod_i\left(1+|z_i|^2/(4R^2)\right)^{-N_\phi/2}\cdot\prod_k\left(1+
|w_k|^2/(4R^2)\right)^{-N/2};\non
\label{nqhole}\eea%
this state is not yet normalized.
Here and below we ``re-use'' indices in distinct products separated by dots $\cdot$. As all quasiholes are at general positions, none are at the south pole, and so the number of flux (determined from the $z_i$ dependence) is $N_\phi=Q(N-1)+n$. For the quasiholes, in view of the single-particle discussion, the function makes sense on the sphere as a function of one $w_k$ (as a section of a bundle over the sphere) with the $N$ $z_i$s and remaining $n-1$ $w_l$s held fixed, provided the total number of flux seen by each quasihole is $N$ \cite{hald83}. This result, which comes simply from counting factors $z_i-w_k$ or $\left(1+|w_k|^2/(4R^2)\right)^{-1/2}$, can be viewed as the analog of the monodromy in the statistics calculations in Ref.\ \cite{read08b}.

In order to calculate adiabatic transport of the quasiholes, we need to know the normalization of these states as functions of $w_k$. This was addressed in the plane years ago; see Refs.\ \cite{halp,stonebook,read08b}. There we need to introduce factors $\prod_{k<l}(w_k-w_l)^{1/Q}$, as well as non-holomorphic factors. Turning this into the form suitable for the sphere (and noting that it agrees with the plane for relatively close quasiaholes, as well as in the limit $N_\phi\to\infty$ with $N_\phi/R^2$ and $n$ fixed), we are led to the functions%
\bea%
\lefteqn{\Psi(w_1,\ldots,w_n;z_1,\ldots,z_N)=}\non\\
&&\prod_{k<l}(w_k-w_l)^{1/Q}\cdot
\prod_{i,k}(z_i-w_k)\cdot\prod_{i<j}(z_i-z_j)^Q\non\\
&&{}\cdot
\prod_i\left(1+|z_i|^2/(4R^2)\right)^{-N_\phi/2}\non\\&&{}\cdot\prod_k\left(1+
|w_k|^2/(4R^2)\right)^{N_\phi^{\rm( qh)}/2},
\label{halpqhole}\eea%
where %
\be%
N_\phi^{\rm(qh)}=-N-(n-1)/Q.\ee%
These now exhibit the monodromy in the $w_k$ that corresponds to fractional statistics \cite{halp}. Moreover, because we introduced factors for each quasihole that change the degree in $w_k$, and which correspond to a ``singular gauge transformation'', the flux (Chern number) for the bundle for each quasihole has changed, to the value $-N_\phi^{\rm(qh)}$. (The sign in the definition is to incorporate the fact that the charge of the quasiholes has the opposite sign to the particles. Note that these are wavefunctions for the particles, and that a coherent state wavepacket, as above, was anti-holomorphic in the parameter. By contrast, the wavefunctions are holomorphic in the $w_k$s which are the parameters for coherent states of quasiholes, not coordinates.) Again, this can also be viewed as a monodromy argument, as the coordinates other than $w_k$ are held fixed, and no adiabatic transport (Berry phase) was calculated so far. There is no need to be alarmed by the fact that this Chern number is in general not an integer, because we have now changed the rules by allowing a section to be not single valued (it is still single valued in the south-pole patch, provided no other quasiholes are present in that region).

Now we turn to normalization and the Berry connection. The main tool is the plasma mapping \cite{laugh}. In the conventions of Ref.\ \cite{read08b}, the particles correspond to charges $Q$, and the quasiholes to impurities of charge $1$ (note that in this point of view, particles and quasiholes are on a similar footing, and have the same sign for these charges). Then there must be a neutralizing background charge density, which is associated with the magnetic field and is clearly uniform for symmetry reasons. The net background charge is clearly $N_b=-(QN+n)$. This background charge density is represented by the non-holomorphic factors in the $z_i$s and $w_k$s, like the Gaussian factors in the plane case. However, the net background charge does not equal $-Q$ times the number of flux $N_\phi$.

To understand this, we are forced to turn to the conformal field theory (CFT) point of view \cite{mr,read08b}. Then the particles and quasiholes (and also the small fixed charges making up the neutralizing background) correspond to fields (operators) in the two-dimensional (2D) CFT. The wavefunctions in the present form are supposed to be conformal blocks for these fields. As such, each field should carry a conformal weight, which for the present case, for a field of charge $q$ (in the plasma normalization) is $q^2/(2Q)$. Thus the particles have conformal weight $Q/2$, and the quasiholes $1/(2Q)$. The conformal weight also determines the spin, and this spin couples to the curvature of the sphere, as described above. Thus, quite generally in CFT, conformal blocks on the sphere will contain non-holomorphic factors that describe the coupling to the curvature. (Usually, CFT is formulated on the flat plane, and such factors are absent.) For the round metric on the sphere as used here, these factors have the same form as those for the magnetic field, or for the neutralizing background charge, and can be deduced in exactly the same way from rotational invariance as we did earlier for LLL wavefunctions. Namely, in CFT, a conformal block for a field with conformal weight $h$ (or spin $s=h$) at $z$ in the plane behaves as $z^{-2h}$ as $z\to\infty$, which controls the net number of factors of $z-z'$ for all other fields at positions $z'$. On the sphere this is then accompanied by the factor%
\be%
\left(1+|z|^2/(4R^2)\right)^h\ee%
for each field. Naturally, the conformal weight also shows up in the bundle point of view when all other fields are held at fixed positions. In both places, the net curvature of the sphere corresponds to a flux of $2s$ when the spin is $s$ (up to a sign depending on how we are defining flux). Note that here we are treating the background charge as discrete charges, but that these produce additional non-holomorphic factors when the continuous limit is taken \cite{mr}.

These considerations lead us now to view the particles as carrying spin $Q/2$, different from the original point of view in which the wavefunctions are scalar functions of the particle coordinates. In addition to the curvature, the neutralizing background is also equivalent to some flux. (More technically, the bundle is a tensor product of the appropriate spin bundle, times one for the
extra flux.) We should have the total flux (or non-holomorphic factor in each $z_i$)%
\bea%
N_\phi&=&-N_b-2s\\
&=&QN+n-Q,\eea
which agrees with the expressions above. The quasiholes interact with the same background charge density, though more weakly by a factor of $Q$, and the same is true for the ``extra'' flux in the bundle point of view. So we should have%
\bea%
-N_\phi^{\rm(qh)}&=&-N_b/Q-2s_{\rm qh}\\
&=&N+n/Q-1/Q,\eea
which also agrees. (Again, these statements correspond to monodromy, but now for the states after the singular gauge transformation, and give the result for $s_{\rm qh}$ in the sense of monodromy.) Then we can conclude that the above functions $\Psi$ are conformal blocks for the fields stated (including the uniform neutralizing background).

The mod-square functions $|\Psi|^2$ can now be viewed as the Boltzmann weight of a plasma with impurities at $w_k$. Charge neutrality should hold locally on scales larger than the screening length (the main point of the preceding three paragraphs was to explain what this ``neutrality'' means). This implies that the functions $\Psi$ are normalized (up to $w_k$-independent factors) when all $w_k$s are separated by distances much greater than the screening length.

Because we now have functions that are normalized (in the region of interest, i.e.\ for well-separated $w_k$s), and also holomorphic in the coordinates $w_k$ up to explicitly-known factors, we can easily calculate the Berry connection and curvature. We do this here for one $w_k$ moving counterclockwise on a small loop that does not enclose any other quasiholes. Because we did a similar calculation earlier for a single particle wavepacket, which was anti-holomorphic in the parameter up to similar non-holomorphic factors, we can simply read off the result. It is proportional to the area of the loop, but it will be convenient to state the result normalized to the surface area of the sphere.
The result can be read off from the number of non-holomorphic factors involving $|w_k|^2$ in the wavefunction, and is simply $N_\phi^{\rm (qh)}$ (with the sign as defined). Note that this result follows once we know that the wavefunctions are normalized, without use of our interpretation about $N_b$ and particle spin. But now using those ideas, as we have identified the effective magnetic flux as $N_b$, a corresponding amount $N_b/Q$ can be subtracted, and the remainder is twice the adiabatic spin $s$ of this quasihole, so $s_{\rm qh}=1/(2Q)$. Thus the monodromy and holonomy results agree. Note that a similar calculation can be done for the quasihole of charge $-q/Q$, $q>0$ (in units of the particle charge), with the result $q^2/(2Q)$. In particular, for a ``real'' hole (missing particle), we obtain spin $Q/2$. For states containing only these ``real'' holes, the fact that they are normalized can be shown directly (when the holes are well separated): the holes are made by applying  destruction field operators to the Laughlin ground state, and the norm-square is essentially the expectation of a product of density operators. We must still use the screening in the plasma to infer that this expectation equals a constant, independent of the positions.

Here we calculated the holonomy for a small closed loop by taking the line integral of the Berry connection, which by Stokes' theorem equals the integral of the curl of the Berry connection inside the loop. If we take the contribution of the outside of the loop instead, we should get the same result. It is worth checking this, because the Berry connection is in a different gauge in the two cases. In particular, without loss of generality, we can apply this to a small loop surrounding the south pole. Then (using the south-pole coordinate patch) the net flux enclosed goes to zero as the loop shrinks to a point. The larger outside of the loop, viewed in the north-pole coordinate patch, encloses the total flux $N_\phi^{\rm (qh)}$ from the Berry connection, and is fractional. As the functions $\Psi$ are not single valued in $w_k$, for the outside we must remember to include also the monodromy when calculating the net Berry phase \cite{read08b}.  The monodromy gives the phase $2\pi (n-1)/Q$ (mod $2\pi$). Hence the net holonomy (Berry phase factor) is $1$ for a vanishing small loop (very much like the argument at the beginning of this paper, but now generalized to include background charge/flux). A similar argument shows that the holonomy for {\em any} loop is independent of which way the loop is viewed.

Now we can generalize to trial wavefunctions that contain a conformal block from some other CFT, in addition to that for the charge part (the Laughlin factor). That is (for a recent discussion see Ref.\ \cite{read08b}; we omit here a label for the distinct blocks in non-Abelian cases, as it does not enter the calculation), %
\bea%
\lefteqn{\Psi(w_1,\ldots,w_n;z_1,\ldots,z_N)=}\\
&&\Psi_{\rm
charge}\cdot\langle\psi(z_1)\cdots\psi(z_N)\tau(w_1)\cdots\tau(w_n) \rangle_{\rm CFT}.\non\eea%
Here the charge part $\Psi_{\rm charge}$ has a form similar to the functions $\Psi$ above, but with the $Q$ replaced by $\nu^{-1}$ ($\nu$ is the filling factor), and the exponent $q=1$ in the factors $z_i-w_k$ replaced by $q_{\rm qh}$. The fields $\psi$ and $\tau$ that appear in the CFT correlator have conformal weights $h_\psi$ and $h_\tau$, respectively. (We consider only one type of quasihole, but the generalization to others will be immediate.) Then on the sphere, the number of flux seen by the electrons is %
\be%
N_\phi=\nu^{-1}(N-1)+nq_{\rm qh}-2h_\psi,\ee%
so that the ``shift'' $S=\nu^{-1}+2h_\psi$ (for $n=0$) is twice the conformal weight of the field representing the particles. We can make similar arguments as in the plasma mapping for the Laughlin states if we assume that the 2D perturbed CFT is in a massive phase \cite{read08b}. Then the wavefunctions are normalized when the quasiholes are well separated. Then a corresponding argument using the total conformal weight for the quasiholes shows that the adiabatic spin of the quasiholes is now%
\be%
s=q_{\rm qh}^2\nu/2+h_\tau,\ee%
which again agrees with the ``monodromy'' result in the conformal blocks. This is the central result of this paper. Note that this quasihole has charge $-q_{\rm qh}\nu$ in units of the particle charge. We emphasize that despite the apparent brevity of the general version of the argument, there is considerable depth behind it, which was explored thoroughly in Ref.\ \cite{read08b}.

We can immediately apply this result to complete the argument about non-unitary CFTs in Ref.\ \cite{read08b}. There the spin and the twist of the quasiparticles were not calculated adiabatically. Now we have shown that the spin $s_\alpha$, and hence the twist $\theta_\alpha=e^{2\pi i s_\alpha}$ for type $\alpha$ are the same as in the underlying CFT (including the charge sector). It then follows by the arguments of Ref.\ \cite{read08b} (which will not be repeated here) that
the quantum dimensions associated with the quasiparticle types are the same as in the CFT. In a non-unitary CFT, some of these may be negative. It can be shown that the presence of negative conformal weights in a rational CFT implies that some quantum dimensions are negative, but it is not known if all non-unitary CFTs contain negative quantum dimensions. Negative quantum dimensions are not acceptable in a topological phase of matter in a conventional quantum theory (i.e. one with positive inner product on Hilbert space). Hence there is an inconsistency arising from the use of any such non-unitary CFT, which indicates that the assumption that the perturbed CFT is in a massive phase breaks down, and this may mean that the trial wavefunctions correspond to a gapless phase or critical point.

Some readers may object that the way the neutralizing background contribution was subtracted in the charge sector is merely a choice and is unconvincing, especially as it leads to an unconventional spin for the particles themselves.  While this objection has some merit, we note that no such subtraction had to be made for the CFT with the charge sector removed, and that the contribution to the spin appears unambiguous. It is feasible to remove the charge part from the theory completely after performing the calculations, and interpret what is left as a topological phase of matter. As the negative quantum dimensions arise in exactly the same way, the last argument still appears to work.

To conclude, we have shown that spin of quasiholes can be calculated by adiabatic transport, and agrees with the monodromy prediction from the trial wavefunctions, which is the conformal weight of the corresponding field in the conformal field theory, when the related 2D field theory is in a massive phase \cite{read08b} (generalizing the plasma argument). This helps rule out the use of non-unitary conformal field theories in the construction of topological phases.

This work was supported by NSF grant no.\ DMR-0706195.

\end{document}